\documentclass{Interspeech}
\usepackage{comment}
\usepackage{tabularx}
\usepackage{booktabs}
\usepackage{pifont}
\usepackage{amsmath,amssymb}
\usepackage{caption}
\usepackage{subcaption}
\usepackage{cite}

\interspeechcameraready 

\title{SpecTokenizer: A Lightweight Streaming Codec in the Compressed Spectrum Domain}

\author[affiliation={1, 2}]{Zixiang}{Wan}
\author[affiliation={1}]{Guochang}{Zhang}
\author[affiliation={1}]{Yifeng}{He}
\author[affiliation={1*}]{Jianqiang}{Wei}
\affiliation{Audio Innovation Technology Department}{Anker Inc}{China}
\affiliation{School of Computer Science}{Beijing University of Posts and Telecommunications}{China}
\email{wzx802@bupt.edu.cn, \{baron.zhang, nolan.he, alex.wei\thanks{* Corresponding author.}\}@anker-in.com}


\begin{document}

\maketitle

\begin{abstract}
    Neural Audio Codecs (NACs) have gained growing attention in recent years as technologies for audio compression and audio representation in speech language models. While mainstream NACs typically require G-level computation and M-level parameters, the performance of lightweight and streaming NACs remains underexplored. This paper proposes SpecTokenizer, a lightweight streaming codec that operates in the compressed spectral domain. Composed solely of alternating CNN and RNN layers, SpecTokenizer achieves greater efficiency and better representational capability through multi-scale modeling in the compressed spectrum domain. At 4 kbps, the proposed SpecTokenizer achieves comparable or superior performance compared to the codec with state-of-the-art lightweight architecture while requiring only 20\% of the computation and 10\% of the parameters. Furthermore, it significantly outperforms the codec when using similar computational and storage resources.
  \end{abstract} 
\section{Introduction}
\label{intro}

Neural Audio Codec (NAC) compresses audio signals into sequences of discrete codes for efficient storage and transmission \cite{guo2025recent, wu-etal-2024-codec, shi2024espnet, wu2024ts3, wu2024codec}. Recently, NAC has gained growing attention as a key technology in Speech Language Modeling (SLM) \cite{yao2025gense, guo2024socodec, ji2024language}. By converting continuous audio signals into discrete codes, Large Language Models (LLMs)—which have achieved great success in text processing—can now be applied to various speech processing tasks \cite{wang2024comparative, wu2024towards, huang2024audiogpt}.

Many high-performance NAC methods have been proposed, optimized for higher audio quality, bitrate efficiency, and lower computational cost \cite{zeghidour2021soundstream, defossez2022high, kumar2024high, du2024funcodec}. However, most of these models have G-level computational complexity and M-level parameter counts, making them unsuitable for deployment on edge hardware. The performance of models with lower computational and parameter scales has not been thoroughly explored. On the other hand, in signal-level tasks like speech enhancement \cite{zhang2022multi}, speech signal improvement \cite{ristea2025icassp} and music source separation \cite{lu2024music}, spectrogram-based models often perform better. Yet, in audio codecs, spectrogram-based codecs often underperform compared to waveform-based codecs. To bridge these gaps, we propose SpecTokenizer.

When NACs are used for audio reconstruction and as tokenizers in SLMs, the following features are particularly important:

\begin{itemize}
    \item \textbf{Audio Quality}: Critical for NAC performance, impacting users and tasks. NACs should faithfully reconstruct audio details without distortion, vital for applications like speech communication and modeling.

    \item \textbf{Bitrate Efficiency}: Refers to transmitting information efficiently per unit of bitrate. A NAC with high bitrate efficiency can achieve comparable or superior audio quality at a lower bitrate.

    \item \textbf{Streaming}: To achieve real-time interaction with simultaneous speaking and responding, codecs need streaming support for low-latency speech encoding and response generation.

    \item \textbf{Single Codebook}: Single-codebook models are favored over multi-codebook models since the latter introduce additional design complexity to the SLM architecture \cite{yang2024uniaudio, du2024cosyvoice, yang2023uniaudio}.

    \item \textbf{Low Computational Complexity}: NACs with low computational complexity enable faster encoding and decoding, reducing computational demands and allocating more resources to SLMs.

    \item \textbf{Low Parameter Count}: NACs with fewer parameters require less storage, reducing memory demands and freeing up resources for edge hardware.

    \item \textbf{Low Token Rate}: Longer sequences can slow down and destabilize LLM training. Therefore, it's preferable to use NAC models with a low token rate to support SLMs.
\end{itemize}

This paper introduces SpecTokenizer, a lightweight streaming codec in the compressed spectral domain—the first neural audio codec with M-level computation and K-level parameters. Composed solely of alternating CNN and RNN layers, SpecTokenizer achieves efficiency and enhanced representational capability through multi-scale modeling in the compressed spectrum domain. SpecTokenizer offers streaming capability, low computational and memory demands, a wide bitrate range, and a single codebook design, while maintaining high audio quality. Under similar computational and storage resources, SpecTokenizer significantly improves audio quality compared to state-of-the-art codecs. Furthermore, after model miniaturization, SpecTokenizer achieves comparable or superior performance using only 20\% of the computation and 10\% of the parameters.
\section{Related work}
\label{related}

In recent years, most well-known neural audio codecs have been based on vector quantization (VQ) networks \cite{van2017neural}, such as SoundStream \cite{zeghidour2021soundstream}, EnCodec \cite{defossez2022high}, and DAC \cite{kumar2024high}. Among them, SoundStream is the first universal codec handling multiple types of audio. EnCodec improves compression rates by integrating a lightweight Transformer language model in the discrete latent space. Building on this, researchers \cite{kumar2024high} have explored quantization dropout, a bitrate scalable technique. They have also shown that periodic inductive bias functions outperform common activation functions in audio signal modeling. These models process audio waveforms directly and are therefore categorized as waveform-based codecs.

In contrast, spectrogram-based codecs focus on processing more intuitive audio spectral features. For example, Lyra \cite{kleijn2021generative} converts audio waveforms into log mel-spectrograms and directly quantizes them into tokens. Due to the irreversibility of mel-spectrograms, Lyra relies on a vocoder to synthesize waveforms. To avoid the inefficiency of heavy vocoders, codecs like TFNet \cite{jiang2022end} and ESC \cite{gu2024esc} use the invertible Short-Time Fourier Transform (STFT) to convert waveforms into complex spectra. This design allows the reconstructed STFT spectra to be losslessly converted back into waveforms via the inverse STFT. FreqCodec \cite{du2024funcodec} introduces depthwise and grouped convolutions into the encoder and decoder. This significantly reduces computational and parameter complexity while maintaining speech quality comparable to state-of-the-art codecs at the time.
\section{Method}
\label{method}

\begin{figure}[ht]
    \centering
    \includegraphics[width=\columnwidth]{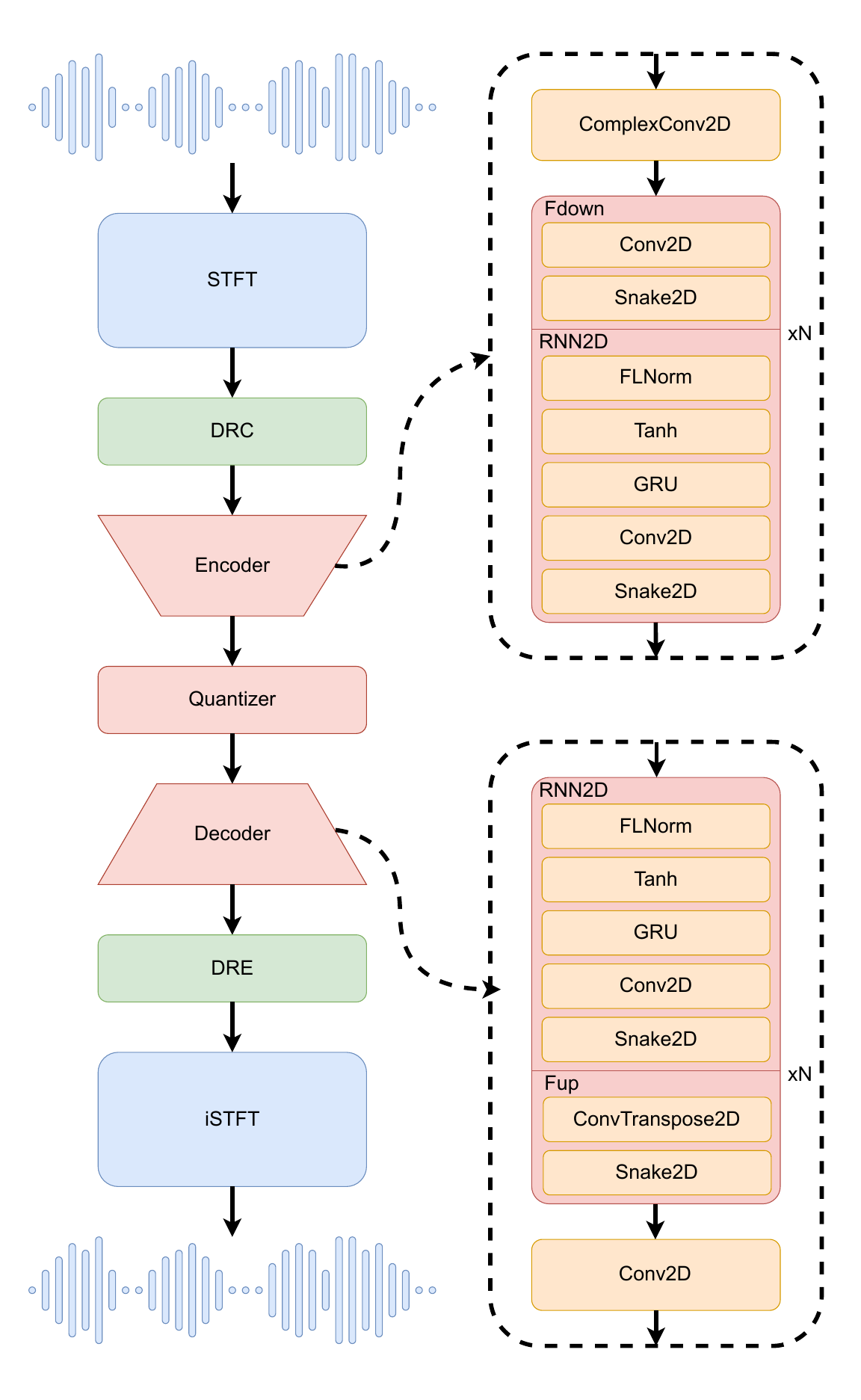}
    \caption{\normalfont
    The architecture of the proposed SpecTokenizer. 
    \textbf{DRC} denotes dynamic range compression, \textbf{DRE} denotes dynamic range expansion,
and \textbf{FLNorm} denotes frame-wise layer normalization.
}
    \label{fig:framework}
\end{figure}

\subsection{Overview}

Figure~\ref{fig:framework} shows the architecture of SpecTokenizer. The speech signal $x$ is transformed into a complex spectrogram $s$ via STFT, followed by \textbf{Dynamic Range Compression (DRC)} to obtain $s_c$. The encoder $E$ extracts latent features $z$ from $s_c$, and the quantization layer $Q$ generates the compressed representation $z_q$. The decoder $G$ reconstructs the compressed spectrum $\hat{s}_c$. After \textbf{Dynamic Range Expansion (DRE)} to get $\hat{s}$, the inverse STFT (iSTFT) is applied to obtain the reconstructed audio signal $\hat{x}$.

\subsection{Spectral Compression}

Experience shows that audio signals typically have extremely high dynamic ranges in amplitude, which poses challenges for directly processing raw spectral data. To address this, we perform \textbf{spectral compression} on the spectrum. The core idea is to apply a compression function to the amplitude while preserving the phase.

In the dynamic range compression stage, given a complex spectrum \(s\), we compress it as:

\begin{equation}
s_c = f( |s| ) \cdot \operatorname{sign}(s), \quad \text{where } f( |s| ) = |s|^{1/p}
\end{equation}

Here, \(|s|\) denotes the magnitude of \(s\), \(\operatorname{sign}(s) = s / |s|\) represents its phase, and \(p > 0\) is the compression coefficient.

In the dynamic range expansion stage, after obtaining the reconstructed compressed spectrum $\hat{s}_c$ from the model, we apply the inverse function $f^{-1}$ to convert the compressed-scale spectrum back into the linear-scale spectrum:

\begin{equation}
\hat{s} = f^{-1}( |\hat{s}_c| ) \cdot \operatorname{sign}( \hat{s}_c ), \quad \text{where } f^{-1}( |s_c| ) = |s_c|^{p}
\end{equation}






\subsection{Encoder and Decoder}

Our model's encoder and decoder architecture is inspired by MTFAA-Net \cite{zhang2022multi}. The encoder $E$ starts with a 2D complex convolution, followed by $N$ alternating frequency down-sampling blocks (\textit{Fdown}) and RNN2D blocks. Each \textit{Fdown} block contains a 2D strided convolution with channel number $C$, kernel size $K$, stride $S$, and the Snake2D activation function. Each RNN2D block includes a \textbf{\underline{F}}rame-wise \textbf{\underline{L}}ayer \textbf{\underline{Norm}}alization \textbf{(FLNorm)}, a Tanh activation function, a GRU layer, a 2D convolution with kernel size and stride set to 1, and the Snake2D activation function \cite{ziyin2020neural}.

RNNs are suitable for tasks with short historical dependencies like audio encoding and decoding, and they have low memory consumption during inference, making them ideal for lightweight edge deployment. The frame-wise layer normalization is specifically designed for RNNs, featuring streaming processing, stable training, and accelerated convergence. It normalizes over the channel and frequency dimensions at each time step:

\begin{equation}
\mu_{t} = \frac{1}{C \times F} \sum_{c=1}^{C} \sum_{f=1}^{F} Y_{c, t, f}
\end{equation}

\begin{equation}
\hat{Y}_{c, t, f} = \frac{Y_{c, t, f} - \mu_{t}}{\sqrt{\sigma^2_{t} + \epsilon}}
\end{equation}

where $Y_{c,t,f}$ is the input tensor; $C$ and $F$ are the channel and frequency dimensions; $\mu_{t}$ and $\sigma^{2}_{t}$ are the mean and variance over $C$ and $F$ at time $t$; and $\epsilon$ is a small constant to prevent division by zero.

The Snake2D activation function captures periodic features in the data. The decoder mirrors the encoder's structure but uses transposed convolutions instead of strided convolutions, with strides opposite to those of the encoder.

\subsection{Discriminator}

We use two discriminators: the Multi-Period Discriminator (MPD) from HiFi-GAN \cite{kong2020hifi}, and the frequency-domain Multi-Band Multi-Scale STFT Discriminator (MBMS-STFT) from DAC \cite{kumar2024high}, which improves phase modeling in practice.

\subsection{Loss Function}

The loss function of SpecTokenizer is written as follows:
\begin{equation}
    \mathcal{L} = \lambda_{\text{rec}}\, \mathcal{L}_{\text{rec}} + \lambda_{\text{adv}}\, \mathcal{L}_{\text{adv}} + \lambda_{\text{feat}}\, \mathcal{L}_{\text{feat}} + \lambda_{\text{cmt}}\, \mathcal{L}_{\text{cmt}}
\end{equation}
where $\lambda_{\text{rec}}$, $\lambda_{\text{adv}}$, $\lambda_{\text{feat}}$, and $\lambda_{\text{cmt}}$ are weight coefficients for the reconstruction loss $\mathcal{L}_{\text{rec}}$, adversarial loss $\mathcal{L}_{\text{adv}}$, feature matching loss $\mathcal{L}_{\text{feat}}$, and commitment loss $\mathcal{L}_{\text{cmt}}$, respectively.

\subsection{Single Large Codebook}

In the single large codebook setting, we initialize the codebook using a clustering method and update it via exponential moving average (EMA) \cite{van2017neural}. To improve codebook utilization, we introduce factorized codebook lookup \cite{kumar2024high} and expiration mechanism\footnote{Due to a bug in the widely used EnCodec code, the expiration mechanism has not been effectively used in many codecs. See \url{https://github.com/facebookresearch/encodec/issues/61}} \cite{dhariwal2020jukebox}. Since the codebook size is much larger than the batch data volume—leading to high repetition in batch-sampled data—we add a data pool module. This module accumulates historical data on a larger scale to better represent the data distribution. 

\section{Experiments}
\label{experiments}

\begin{table}[t]
    \centering
    \fontsize{8}{9}\selectfont
    \setlength\tabcolsep{1pt}
    \caption{Comparison of SpecTokenizer and baseline codecs.
    }
    \label{tab:baselines}
    \begin{tabularx}{\linewidth}{@{}l c c >{\centering\arraybackslash}X@{}}
        \toprule
        \textbf{Codec} & \textbf{Streaming} & \textbf{VQ Type} & \textbf{Domain} \\
        \midrule
        SpecTokenizer (Ours) & \ding{51} & RVQ / Single & Compressed Complex \\
        EnCodec \cite{defossez2022high} & \ding{51} & RVQ & Waveform \\
        HiFi-Codec \cite{yang2023hifi} & \ding{55} & GRVQ & Waveform \\
        DAC \cite{kumar2024high} & \ding{55} & RVQ & Waveform \\
        FreqCodec \cite{du2024funcodec} & \ding{51} & RVQ & Magnitude, Phase \\
        SpeechTokenizer \cite{zhang2023speechtokenizer} & \ding{55} & RVQ & Waveform \\
        WavTokenizer \cite{ji2024wavtokenizer} & \ding{55} & Single & Waveform \\
        \bottomrule
    \end{tabularx}
\end{table}
\begin{table*}[t]
    \centering
    \fontsize{9}{10}\selectfont
    \setlength\tabcolsep{7pt}
    \captionsetup{justification=centering} 
    \caption{Comparison of different codec models at various bitrates. (a) Rows 1--6; (b) Rows 7--11; (c) Rows 12--17. \\ The SpecTokenizer$^*$ indicates SpecTokenizer with a single large codebook of size 32k.}
    \label{tab:mainresults}
    \begin{tabularx}{\linewidth}{@{}l c c c c c c c c@{}}
        \toprule
        \textbf{Codec} & \textbf{Bitrate (kbps)} & \textbf{WER(\%)}$\downarrow$ & \textbf{PESQ}$\uparrow$ & \textbf{STOI}$\uparrow$ & \textbf{SDR}$\uparrow$ & \textbf{MelLoss}$\downarrow$ & \textbf{UTMOS}$\uparrow$ & \textbf{XLSR-MOS}$\uparrow$ \\
        \midrule
        Ground Truth             & - & 2.87  & -     & -      & -      & -      & 4.04 & 4.08 \\
        EnCodec                  & 12  & 3.28  & 3.07  & 0.943  & 10.14  & 0.929  & 3.51 & 3.97 \\
        EnCodec                  & 6   & 3.63  & 2.47  & 0.909  & 7.48   & 1.033  & 3.04 & 3.60 \\
        DAC                      & 6   & 3.13  & 3.45  & 0.948  & 1.03   & 0.566  & 3.97 & 4.05 \\
        FreqCodec                & 6   & 3.39  & 2.63  & 0.821  & 3.98   & 0.885  & 3.84 & 3.97 \\
        SpecTokenizer            & 6   & \textbf{2.74}  & \textbf{3.47}  & \textbf{0.957}  & \textbf{10.47}  & \textbf{0.531}  & \textbf{4.01} & \textbf{4.06} \\
        \midrule
        HiFi-Codec               & 4   & 5.62  & 2.05  & 0.866  & 3.50   & 1.652  & 3.57 & 3.70 \\
        SpeechTokenizer          & 4   & 4.34  & 1.98  & 0.843  & 1.52   & 0.842  & 3.87 & 3.94 \\
        DAC                      & 4   & 3.55  & 2.47  & 0.902  & 0.11   & 0.719  & 3.43 & 3.77 \\
        FreqCodec                & 4   & 3.44  & 2.46  & 0.814  & 3.14   & 0.917  & 3.79 & 3.92 \\
        SpecTokenizer            & 4   & \textbf{2.89}  & \textbf{3.04}  & \textbf{0.935}  & \textbf{8.36}   & \textbf{0.601}  & \textbf{3.98} & \textbf{4.04} \\
        \midrule
        WavTokenizer             & 0.5 & 19.97 & 1.39  & 0.740  & -9.98  & 1.112  & \textbf{3.57} & \textbf{3.87} \\
        SpeechTokenizer          & 0.5 & \textbf{10.63} & 1.16  & 0.572  & -14.58 & 1.813  & 1.26 & 2.83 \\
        DAC                      & 0.5 & 78.82 & 1.08  & 0.583  & -7.93  & 1.946  & 1.25 & 1.35 \\
        FreqCodec                & 0.5 & 22.57 & 1.48  & 0.718  & \textbf{-2.02}  & 1.246  & 2.97 & 3.14 \\
        SpecTokenizer            & 0.5 & 20.32 & 1.44  & 0.755  & -3.06  & 1.009  & 2.90 & 3.19 \\
        SpecTokenizer$^*$        & 0.5 & 14.82 & \textbf{1.51}  & \textbf{0.767}  & -2.42  & \textbf{0.957}  & 3.10 & 3.47 \\
        \bottomrule
    \end{tabularx}
\end{table*}

\begin{table}[t]
    \centering
    \fontsize{8}{9}\selectfont
    \setlength\tabcolsep{4pt} 
    \caption{Comparison of cost and performance of different codec models at 4 kbps. For the mini version of SpecTokenizer, we set $C = [16, 24, 32, 64, 96]$.}
    \label{tab:mini}
    \begin{tabularx}{\linewidth}{@{}l c c c c@{}}
        \toprule
        \textbf{Codec} & $\textbf{FLOPs (G)}$ & $\textbf{Params (M)}$ & $\textbf{PESQ}$ & $\textbf{SDR}$ \\
        \midrule
        DAC & 55.66 & 74.06 & 2.47 & 0.11 \\
        FreqCodec & 2.18 & 4.50 & 2.46 & 3.14 \\
        HiFi-Codec & 53.62 & 62.56 & 2.05 & 3.50 \\
        SpeechTokenizer & 17.09 & 103.68 & 1.98 & 1.52 \\
        \midrule
        SpecTokenizer (mini) & \textbf{0.43} & \textbf{0.45} & 2.51 & 6.02 \\
        SpecTokenizer (base) & 56.40 & 70.61 & \textbf{3.04} & \textbf{8.36} \\
        \bottomrule
    \end{tabularx}
\end{table}


\begin{table}[t]
    \centering
    \fontsize{8}{9}\selectfont
    \setlength\tabcolsep{4pt}
    \caption{Results of the ablation study on our proposed codec. }
    \label{tab:ab_backbone}
    \begin{tabularx}{\linewidth}{@{}l c c c c c@{}}
        \toprule
        \textbf{Codec} & $\textbf{WER(\%)} $ & $\textbf{PESQ}$ & $\textbf{SDR}$ & $\textbf{MelLoss}$ & $\textbf{UTMOS}$ \\
        \midrule
        SpecTokenizer & \textbf{2.74} & \textbf{3.47} & \textbf{10.47} & \textbf{0.531} & \textbf{4.01} \\
        \quad - DRC\&DRE & 3.11 & 3.39 & 10.26 & 0.775 & 3.95 \\
        \quad - FLNorm & 2.87 & 3.39 & 10.24 & 0.550 & 3.96 \\
        \quad - Snake2D & 2.81 & 3.15 & 8.88 & 0.577 & 3.95 \\
        \quad - Multi-Scale & 3.19 & 3.31 & 9.69 & 0.594 & 3.84 \\
        \bottomrule
    \end{tabularx}
\end{table}
\begin{table}[t]
    \centering
    \fontsize{8}{9}\selectfont
    \setlength\tabcolsep{4pt} 
    \caption{Results of the ablation study on the improvements to the single large codebook setup of SpecTokenizer.}
    \label{tab:ab_slc}
    \begin{tabularx}{\linewidth}{@{}l c c c c@{}}
        \toprule
        \textbf{Codec} & $\textbf{WER(\%)}$ & $\textbf{PESQ}$ & $\textbf{UTMOS}$ & $\textbf{Utilization Rate} (\%)$ \\
        \midrule
        SpecTokenizer & 10.94 & 1.63 & 3.13 & 62.0 \\
        \quad + expirecode & 9.76 & 1.65 & 3.25 & 93.4 \\
        \quad \quad + datapool & \textbf{8.81} & \textbf{1.72} & \textbf{3.48} & \textbf{94.8} \\
        \bottomrule
    \end{tabularx}
\end{table}

\subsection{Dataset}

We used LibriTTS \cite{zen2019libritts} as the training dataset, containing approximately 585 hours of English speech. For evaluation, we used the datasets specified by the Codec-SUPERB challenge at SLT 2024 \cite{wu2024codec}. Codebook utilization, UTMOS \cite{saeki2022utmos} and XLSR-MOS \cite{tamm2023analysis} were evaluated using the LibriTTS test set.

\subsection{Baselines}

We selected state-of-the-art (SOTA) codec models as baselines for SpecTokenizer. To ensure fair comparisons, we employed the official weights provided by the EnCodec \cite{defossez2022high}, HiFi-Codec \cite{yang2023hifi}, DAC \cite{kumar2024high}, FreqCodec \cite{du2024funcodec}, SpeechTokenizer \cite{zhang2023speechtokenizer}, and WavTokenizer \cite{ji2024wavtokenizer}. The differences between SpecTokenizer and these models are shown in Table~\ref{tab:baselines}.

\subsection{Evaluation Metrics}

To comprehensively evaluate our codec, we followed the methods proposed by the Codec-SUPERB challenge at SLT 2024 \cite{wu2024codec}, analyzing the quality of reconstructed audio from both application-level and signal-level perspectives. At the application level, we focused on Automatic Speech Recognition (ASR). At the signal level, we used objective metrics such as PESQ, STOI, SDR, and Mel spectrum loss to quantify audio quality, ensuring the model's performance in preserving speech content and signal fidelity. Additionally, we included codebook utilization, UTMOS and XLSR-MOS.

\subsection{Implementation Details}

Our model accepts audio input at 16 kHz. For spectral compression, the coefficient $p$ is set to 2. For the encoder and decoder, we set the number of blocks $N = 4$, the number of channels $C = [192, 256, 384, 800, 1280]$, convolution kernel sizes $K = [(3, 4), (3, 4), (3, 4), (3, 4)]$, and strides $S = [(1, 4), (1, 4), (1, 4), (2, 4)]$. Each element of $K$ and $S$ corresponds to the time and frequency dimensions, respectively. Downsampling results in a frame rate of 50 frames per second. For the quantizer, we adopt a Residual Vector Quantizer (RVQ) composed of 12 codebooks of size 1024, with dropout set to 0.5. For the MPD and MBMS-STFT discriminators, we maintain the same structure as in DAC. In the single large codebook setting, we set the number of clustering iterations to 50, the EMA decay to 0.99, the expiration threshold to 0.01, and the data pool size to $\max(32768,\ \text{codebook size})$. We train a separate SpecTokenizer model for each bitrate from scratch.

\subsection{Main Results}

\subsubsection{Results at 6 kbps}

Table~\ref{tab:mainresults}~(a) shows the results at a bitrate of $6\,\mathrm{kbps}$. SpecTokenizer achieves the best performance across WER, PESQ, STOI, SDR, MelLoss, UTMOS and XLSR-MOS metrics, with SDR significantly higher than other codecs. This indicates that at $6\,\mathrm{kbps}$, the speech reconstructed by SpecTokenizer has excellent intelligibility, fidelity, and naturalness, closely resembling the original speech.

\subsubsection{Results at 4 kbps}

Table~\ref{tab:mainresults}~(b) shows the results at $4\,\mathrm{kbps}$. We observe the following: (1) SpecTokenizer still performs best on all metrics, with SDR significantly higher than other codecs. Compared to $6\,\mathrm{kbps}$, the advantages of SpecTokenizer over other codecs at $4\,\mathrm{kbps}$ are more pronounced. (2) When the bitrate decreases from $6\,\mathrm{kbps}$ to $4\,\mathrm{kbps}$, DAC's performance drops significantly, while FreqCodec and SpecTokenizer maintain relatively good performance.

\subsubsection{Results at 0.5 kbps}

Table~\ref{tab:mainresults}~(c) shows the results at $0.5\,\mathrm{kbps}$. We observe the following: (1) Among codecs using RVQ, when the bitrate decreases to $0.5\,\mathrm{kbps}$, DAC and SpeechTokenizer's performance drops significantly, while FreqCodec and SpecTokenizer still maintain relatively good performance. (2) Under the single large codebook setting, SpecTokenizer achieves the best PESQ, STOI, and MelLoss metrics, and ranks second on WER, SDR, UTMOS and XLSR-MOS. (3) SpeechTokenizer performs poorly on PESQ, STOI, SDR, UTMOS and XLSR-MOS, but relatively well on WER, possibly because it uses semantic distillation during training, allowing the first codebook's tokens to encode sufficient content information. (4) WavTokenizer performs well on UTMOS and XLSR-MOS but poorly on PESQ, SDR, and MelLoss, indicating that while the reconstructed speech is natural, its fidelity is poor.

\subsubsection{Results of Model Miniaturization}

We conducted model miniaturization experiments at $4\,\mathrm{kbps}$. As shown in Table~\ref{tab:mini}, the miniaturized SpecTokenizer performs comparably or better than FreqCodec while using only 20\% of the computational complexity and 10\% of the parameters. Compared to DAC, SpecTokenizer achieves similar or superior performance using only 0.8\% of DAC's computational complexity and 0.6\% of its parameters. This demonstrates the advantages of spectrogram-based models: they require significantly less computational complexity and fewer parameters than wavform-based models because modeling in the frequency domain is easier, as confirmed by other studies \cite{du2024funcodec, jiang2024mdctcodec}.

\subsection{Ablation Studies}

To further investigate the role of each component in our model, we conducted ablation studies on the architecture by examining the effects of: not performing spectral compression; replacing FLNorm with BatchNorm; replacing Snake2D with ReLU; and eliminating multi-scale modeling by keeping RNN2D only at the largest scale. The experimental results are shown in Table~\ref{tab:ab_backbone}. We found that removing any component reduced model performance to varying degrees. Specifically, replacing Snake2D caused the most severe drop in PESQ, indicating that Snake2D significantly contributes to audio reconstruction quality.

To investigate the effects of each improvement in the single codebook setup, we conducted experiments with a codebook size of 32k, frame rate of 50\,fps, and bitrate of 750\,bps. Table~\ref{tab:ab_slc} shows the effects of each improvement in the single codebook setup of SpeechTokenizer. These improvements are also applicable in other settings.
\section{Conclusion}
\label{Conclusion}
This paper introduces SpecTokenizer, which is the first attempt to develop a neural audio codec (NAC) model with M-level computation and K-level parameters. It features streaming capability, low computational demand, low memory usage, wide bitrate range, and a single codebook design. Compared to the state-of-the-art lightweight codec FreqCodec, SpecTokenizer achieves comparable or better performance using only 20\% of the computation and 10\% of the parameters, and performs significantly better under similar computational and storage conditions. SpecTokenizer opens up a new direction for lightweight and efficient streaming NAC. 


\bibliographystyle{IEEEtran}
\bibliography{mybib}

\end{document}